\documentclass{icrc2009}

\usepackage{graphicx}   
\usepackage[font=footnotesize]{subfig} 
\usepackage{fixltx2e}
\usepackage{url}

\newcommand{\shorttitle}[1]%
{\markboth{Proceedings of the 31\MakeLowercase{$^{st}$} ICRC, {\L}\'{o}d\'{z} 2009}{#1} }

\begin{document}
\title{The Cosmic Ray Energy Spectrum Measured with KASCADE-Grande}

\author{
\IEEEauthorblockN{
A.~Haungs\IEEEauthorrefmark{1}, 
W.D.~Apel\IEEEauthorrefmark{1},
J.C.~Arteaga\IEEEauthorrefmark{2}$^,$\IEEEauthorrefmark{11},
F.~Badea\IEEEauthorrefmark{1},
K.~Bekk\IEEEauthorrefmark{1}, 
M.~Bertaina\IEEEauthorrefmark{3},
J.~Bl\"umer\IEEEauthorrefmark{1}$^,$\IEEEauthorrefmark{2},
}  \\ \vspace{-2.7ex}
\IEEEauthorblockN{
H.~Bozdog\IEEEauthorrefmark{1}
I.M.~Brancus\IEEEauthorrefmark{4},
M.~Br\"uggemann\IEEEauthorrefmark{5},
P.~Buchholz\IEEEauthorrefmark{5},
E.~Cantoni\IEEEauthorrefmark{3}$^,$\IEEEauthorrefmark{6},
A.~Chiavassa\IEEEauthorrefmark{3},
} \\ \vspace{-2.7ex}
\IEEEauthorblockN{
F.~Cossavella\IEEEauthorrefmark{2}, 
K.~Daumiller\IEEEauthorrefmark{1}, 
V.~de Souza\IEEEauthorrefmark{2}$^,$\IEEEauthorrefmark{12}, 
F.~Di~Pierro\IEEEauthorrefmark{3},
P.~Doll\IEEEauthorrefmark{1}, 
R.~Engel\IEEEauthorrefmark{1},
J.~Engler\IEEEauthorrefmark{1}, 
} \\ \vspace{-2.7ex}
\IEEEauthorblockN{
M.~Finger\IEEEauthorrefmark{1}, 
D.~Fuhrmann\IEEEauthorrefmark{7},
P.L.~Ghia\IEEEauthorrefmark{6},
H.J.~Gils\IEEEauthorrefmark{1},
R.~Glasstetter\IEEEauthorrefmark{7}, 
C.~Grupen\IEEEauthorrefmark{5},
D.~Heck\IEEEauthorrefmark{1}, 
} \\ \vspace{-2.7ex}
\IEEEauthorblockN{
J.R.~H\"orandel\IEEEauthorrefmark{2}$^,$\IEEEauthorrefmark{13}, 
T.~Huege\IEEEauthorrefmark{1}, 
P.G.~Isar\IEEEauthorrefmark{1}, 
K.-H.~Kampert\IEEEauthorrefmark{7},
D.~Kang\IEEEauthorrefmark{2}, 
D.~Kickelbick\IEEEauthorrefmark{5},
} \\ \vspace{-2.7ex}
\IEEEauthorblockN{
H.O.~Klages\IEEEauthorrefmark{1}, 
P.~{\L}uczak\IEEEauthorrefmark{8}, 
H.J.~Mathes\IEEEauthorrefmark{1}, 
H.J.~Mayer\IEEEauthorrefmark{1}, 
J.~Milke\IEEEauthorrefmark{1}, 
B.~Mitrica\IEEEauthorrefmark{4},
C.~Morello\IEEEauthorrefmark{6},
} \\ \vspace{-2.7ex}
\IEEEauthorblockN{
G.~Navarra\IEEEauthorrefmark{3}$^,$\IEEEauthorrefmark{15},
S.~Nehls\IEEEauthorrefmark{1},
J.~Oehlschl\"ager\IEEEauthorrefmark{1}, 
S.~Ostapchenko\IEEEauthorrefmark{1}$^,$\IEEEauthorrefmark{14}, 
S.~Over\IEEEauthorrefmark{5},
M.~Petcu\IEEEauthorrefmark{4}, 
T.~Pierog\IEEEauthorrefmark{1}, 
} \\ \vspace{-2.7ex}
\IEEEauthorblockN{
H.~Rebel\IEEEauthorrefmark{1}, 
M.~Roth\IEEEauthorrefmark{1}, 
H.~Schieler\IEEEauthorrefmark{1}, 
F.~Schr\"oder\IEEEauthorrefmark{1}, 
O.~Sima\IEEEauthorrefmark{9}, 
M.~St\"umpert\IEEEauthorrefmark{2}, 
G.~Toma\IEEEauthorrefmark{4}, 
} \\ \vspace{-2.7ex}
\IEEEauthorblockN{
G.C.~Trinchero\IEEEauthorrefmark{6},
H.~Ulrich\IEEEauthorrefmark{1},
A.~Weindl\IEEEauthorrefmark{1},
J.~Wochele\IEEEauthorrefmark{1}, 
M.~Wommer\IEEEauthorrefmark{1}, 
J.~Zabierowski\IEEEauthorrefmark{8}
} 
\IEEEauthorblockA{\IEEEauthorrefmark{1}Institut f\"ur Kernphysik, Karlsruher Institut f\"ur Technologie, 76021~Karlsruhe, Germany}
\IEEEauthorblockA{\IEEEauthorrefmark{2}Institut f\"ur Experimentelle Kernphysik, Karlsruher Institut f\"ur Technologie, 76021 Karlsruhe, Germany}
\IEEEauthorblockA{\IEEEauthorrefmark{3}Dipartimento di Fisica Generale dell'Universit{\`a},
10125 Torino, Italy}
\IEEEauthorblockA{\IEEEauthorrefmark{4}National Institute of Physics and Nuclear Engineering,
7690~Bucharest, Romania}
\IEEEauthorblockA{\IEEEauthorrefmark{5}Fachbereich Physik, Universit\"at Siegen, 57068 Siegen, 
Germany}
\IEEEauthorblockA{\IEEEauthorrefmark{6}Istituto di Fisica dello Spazio Interplanetario, INAF, 
10133 Torino, Italy}
\IEEEauthorblockA{\IEEEauthorrefmark{7}Fachbereich Physik, Universit\"at Wuppertal, 42097
Wuppertal, Germany}
\IEEEauthorblockA{\IEEEauthorrefmark{8}Soltan Institute for Nuclear Studies, 90950~Lodz, 
Poland}
\IEEEauthorblockA{\IEEEauthorrefmark{9}Department of Physics, University of Bucharest, 
76900~Bucharest, Romania}
\IEEEauthorblockA{\small \IEEEauthorrefmark{11}now at: Universidad Michoacana, Morelia, Mexico}
\IEEEauthorblockA{\small \IEEEauthorrefmark{12}now at: Universidade de S$\tilde{a}$o Paulo, Instituto de F\'{\i}sica de S$\tilde{a}$o Carlos, Brasil}
\IEEEauthorblockA{\small \IEEEauthorrefmark{13}now at: Dept. of Astrophysics, Radboud University Nijmegen, The Netherlands}
\IEEEauthorblockA{\small \IEEEauthorrefmark{14}now at: University of Trondheim, Norway}
\IEEEauthorblockA{\bf \IEEEauthorrefmark{15}This article is dedicated to Gianni Navarra, a principle investigator of KASCADE-Grande and an \\outstanding cosmic ray scientist, who passed away much too soon on August 24, 2009.}
}

\shorttitle{A.~Haungs et al. - The KASCADE-Grande Experiment}
\maketitle

\begin{abstract}
KASCADE-Grande is a multi-detector experiment at Forschungszentrum Karlsruhe, Germany for measuring 
extensive air showers in the primary energy range of 100 TeV to 1 EeV. 
This contribution attempts to provide a synopsis of the current results 
of the experiment. In particular, the reconstruction of the all-particle energy spectrum 
in the range of 10 PeV to 1 EeV based on four different methods with partly 
different sources of systematic uncertainties is presented. 
Since the calibration of the observables in terms of the primary energy and mass depends on 
Monte Carlo simulations, we compare the results of various methods applied to the same sample 
of measured data. In addition, first investigations on the elemental composition 
of the cosmic particles as well as on tests of hadronic interaction 
models underlying the analyses are discussed.  
\end{abstract}


\section{Introduction}

The main goal of experimental cosmic ray research is the measurement of the primary
energy spectrum and the elemental composition. 
That comprises important hints to understand the origin, acceleration and
propagation of energetic cosmic particles. 
This task can be done directly or indirectly, depending
on the energy of the primary particle. At high energies, above $10^{15}\,$eV 
the energy spectrum must be determined indirectly
from the measured properties of extensive air showers (EAS) that cosmic rays 
induce in the Earth's atmosphere. 
 
Depending on the experimental apparatus and the 
detection technique, different sets of EAS observables are available to estimate 
the energy of the primary cosmic ray~\cite{Haungs}. In ground arrays the total number 
of charged particles in the shower and the corresponding density at observation level 
are  commonly employed~\cite{Wat}. However, the muon content of EAS can play also 
an important role. One reason in favor of this observable is that
in an air shower muons undergo fewer atmospheric interactions than the electromagnetic 
or hadronic components (dominant component for vertical EAS) and exhibit 
in consequence less fluctuation compared to the more abundant electromagnetic component.

In KASCADE-Grande both components, the muon and the electromagnetic are measured and both, 
together with their correlation on a single-event-basis are used to derive the energy 
spectrum and composition of cosmic rays in the range from $10^{16}$ to $10^{18}\,$eV. 
\begin{figure}[ht]
\vspace*{0.05cm}
\begin{center}
\includegraphics [width=0.39\textwidth]{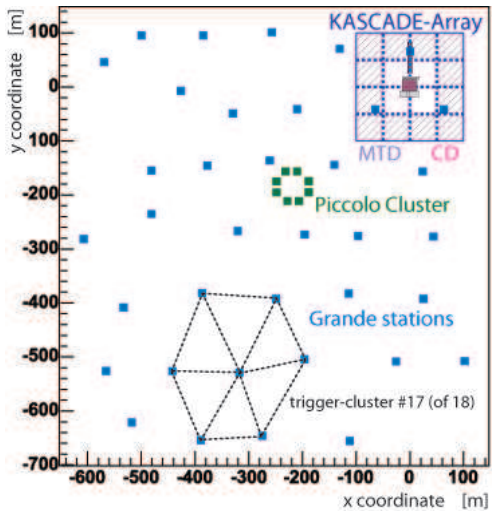}
\end{center}
\vspace*{-0.1cm}
\caption{Layout of the KASCADE-Grande experiment:
The original KASCADE, the distribution of the 37 stations of the Grande
array, and the small Piccolo cluster for fast trigger purposes are shown. 
The outer 12 clusters of the KASCADE array consist of 
$\mu$- and $e/\gamma$-detectors, the inner 4 clusters of 
$e/\gamma$-detectors only.}
\label{fig1}
\vspace*{-0.01cm}
\end{figure}

\section{KASCADE-Grande}

Main parts of the experiment are the Grande array spread over an area of $700 \times 700\,$m$^2$, 
the original KASCADE array covering $200 \times 200\,$m$^2$ with unshielded and shielded 
detectors, and additional muon tracking devices. This multi-detector system allows us to 
investigate the energy spectrum, composition, and anisotropies of cosmic rays in the energy 
range up to $1\,$EeV. The estimation of energy and mass of the primary particles is based 
on the combined investigation of the charged particle, the electron, and the muon components 
measured by the detector arrays of Grande and KASCADE. 

The multi-detector experiment KASCADE~\cite{kascade}
(located at 49.1$^\circ$N, 8.4$^\circ$E, 110$\,$m$\,$a.s.l.)
was extended to KASCADE-Grande 
in 2003 by installing a large array of 37 stations consisting 
of 10$\,$m$^2$ scintillation detectors each (fig.~\ref{fig1}).  
KASCADE-Grande~\cite{Chia09} provides a sensitive area of $\approx 0.4\,$km$^2$
and operates jointly with the existing KASCADE detectors.
The joint measurements with the KASCADE muon tracking devices are 
ensured by an additional cluster (Piccolo) 
close to the center of KASCADE-Grande for fast trigger purposes. 
\begin{figure}[ht]
\vspace*{0.05cm}
\begin{center}
\includegraphics [width=0.42\textwidth]{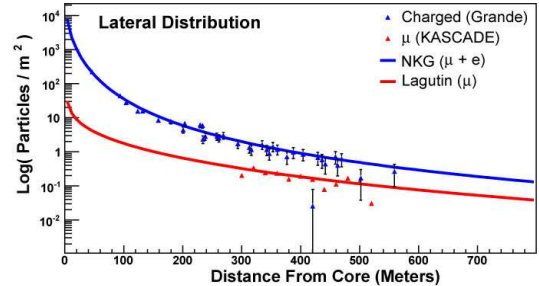}
\end{center}
\vspace*{-0.1cm}
\caption{Example of a typical KASCADE-Grande event. Displayed are the measured densities 
per station for charged particles of the Grande array and averaged over stations in $20\,$m rings 
for the KASCADE muon array, respectively and the corresponding fits of the lateral 
distribution functions.}
\label{figexa}
\vspace*{-0.01cm}
\end{figure}

While the Grande detectors are sensitive to charged particles, 
the KASCADE array detectors measure the electromagnetic 
component and the muonic component separately. 
The muon detectors enable to reconstruct 
the total number of muons on an event-by-event basis
also for Grande triggered events. 
The Muon Tracking Detector (MTD)~\cite{MTD} registers muons above 
an energy threshold of $800\,$MeV. 
At the MTD the directions of muon tracks in EAS are measured with an excellent angular
resolution of $\approx 0.35^{\circ}$. These directional data allow to 
investigate the longitudinal development of the muonic component in showers 
which is used as a signature of the development of the hadronic EAS core.
 
\section{Reconstruction}

Basic shower observables like the core position, angle-of-incidence, 
and total number of charged particles are provided by 
the measurements of the Grande stations. 
By means of Monte Carlo simulations a core position resolution
of $\approx5\,$m, a direction resolution of 
$\approx0.7^\circ$, and a resolution of the total particle number 
in the showers of $\approx15$\% is expected~\cite{dipierro}.  
The total number of muons ($N_\mu$ resolution $\approx25$\%) 
is calculated using the core position determined by the Grande 
array and the muon densities measured by the KASCADE muon 
array detectors~\cite{fuhrmann}.
Fig.~\ref{figexa} shows a typical event of KASCADE-Grande and its 
reconstruction; i.e.~the functions  fitted to the measured lateral 
particle density distributions.  
Full efficiency for triggering and reconstruction of air-showers is reached 
at primary energy of $\approx\:2\:\cdot\:10^{16}\,$eV, slightly depending on
the cuts needed for the reconstruction of the different observables. 
\begin{figure}[ht]
\vspace*{0.05cm}
\begin{center}
\includegraphics [width=0.48\textwidth]{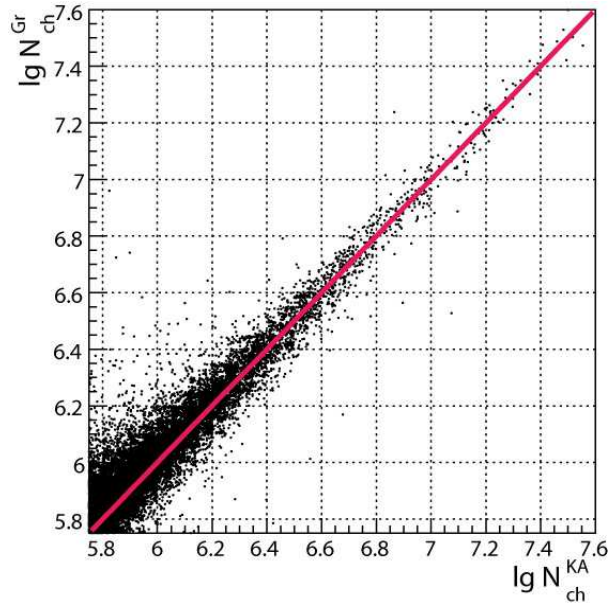}
\end{center}
\vspace*{-0.1cm}
\caption{Scatter plot of the shower sizes (charged particles)
reconstructed by KASCADE (x-axis) and Grande
(y-axis).}
\label{figfed}
\vspace*{-0.01cm}
\end{figure}

For a subsample of events collected by the Grande array it is possible to 
compare on an event-by-event basis the two independent reconstructions of KASCADE and
Grande. This provides the unique opportunity of evaluating the reconstruction 
accuracies of the Grande array by a direct comparison with an independent experiment.
Since the KASCADE array is much more dense, the values of reconstructed observables 
from KASCADE are taken as reference. The subsample is obtained accordingly to
the following selection criteria: maximum energy deposit
in one of the Grande stations close to KASCADE; core position within a circle of 
$90\,$m radius from the KASCADE center; zenith angle less than 40$^\circ$. 
By means of such a comparison (see fig.~\ref{figfed})
the Grande reconstruction accuracies are found to be
for the shower size: systematic uncertainty $\le 5$\%, statistical inaccuracy $\le
15$\%; for arrival direction: $\sigma \approx 0.6^\circ$;
for the core position: $\sigma \approx 5\,$m. 
All of them are in good accordance with the resolutions obtained from simulations.

\section{All-particle energy spectrum}

\subsection{Strategy}

Applying different methods to the same data sample has advantages in various 
aspects: one would expect the same result for the energy spectrum by all 
methods provided the measurements are accurate enough, when the reconstructions work 
without failures, and when the Monte Carlo simulations describe correctly the shower 
development. 
But, the fact that the various observables exhibit substantial differences in their  
composition sensitivity hampers a straightforward analysis.  
However, investigating results of different methods can be used to 
\begin{itemize}
\item cross-check the measurements by different sub-detectors;
\item cross-check the reconstruction procedures;  
\item cross-check the influence of systematic uncertainties;
\item test the sensitivity of the observables to the elemental composition;
\item test the validity of hadronic interaction models underlying the simulations.
\end{itemize}
\begin{figure}[ht]
\vspace*{0.05cm}
\begin{center}
\includegraphics [width=0.49\textwidth]{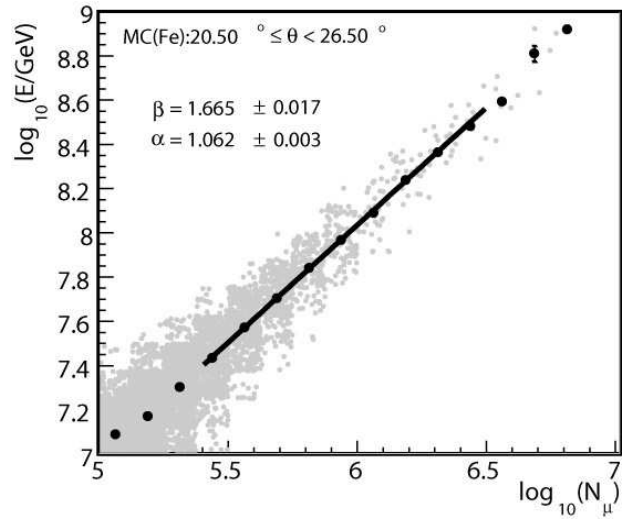}
\end{center}
\vspace*{-0.1cm}
\caption{Mean distribution of true energy vs. muon number 
  for iron nuclei and $\theta = 20.5^{\circ} - 26.5^{\circ}$ calculated with
  MC simulations. The fit with formula  $\mbox{E}[\mbox{GeV}] = \beta \cdot
  N_{\mu}^{\alpha}$ is shown.}
\label{figjuan}
\vspace*{-0.01cm}
\end{figure}

\subsection{Methods}

The estimation of the all-particle energy spectrum is presently based on four different methods
using different observables of KASCADE-Grande:
\begin{itemize}
\item $N_{ch}$-method: The reconstructed charged particle shower size per individual event 
is corrected for attenuation in the atmosphere by the constant intensity cut method and calibrated by Monte Carlo
simulations under the assumption of a  dependence $E_0\propto N_{ch}^{\alpha_{ch}}$ and a particular primary 
composition~\cite{kang}.
\item $N_{\mu}$-method: The reconstructed muon shower size per individual event 
is corrected for attenuation and calibrated by Monte Carlo
simulations under the assumption of a  dependence $E_0\propto N_{\mu}^{\alpha_\mu}$ 
(see example in fig.~\ref{figjuan}) and a particular primary composition~\cite{arteaga}.
\item $N_{ch}-N_\mu$-method: This method combines the information provided by the two observables. 
By help of Monte Carlo simulations a formula is obtained to calculate the 
primary energy per individual shower on basis of $N_{ch}$ and $N_\mu$.  
The formula takes into account the mass sensitivity in order to minimize the composition 
dependence. The attenuation is corrected for by deriving the formula for different zenith 
angle intervals independently and combining the energy spectrum afterwards~\cite{bertaina}.
\item $S(500)$-method: The reconstructed particle density at the specific distance to the shower 
axis of $500\,$m per individual event is corrected for attenuation and calibrated by Monte Carlo 
simulations under the assumption of a dependence $E_0\propto S(500)^{\alpha_{S(500)}}$. 
The distance of $500\,$m  is chosen to have a minimum influence of the primary 
particle type, therefore a smaller dependence on primary composition is expected~\cite{toma}.
\end{itemize}
\begin{figure}[ht]
\vspace*{0.05cm}
\begin{center}
\includegraphics [width=0.49\textwidth]{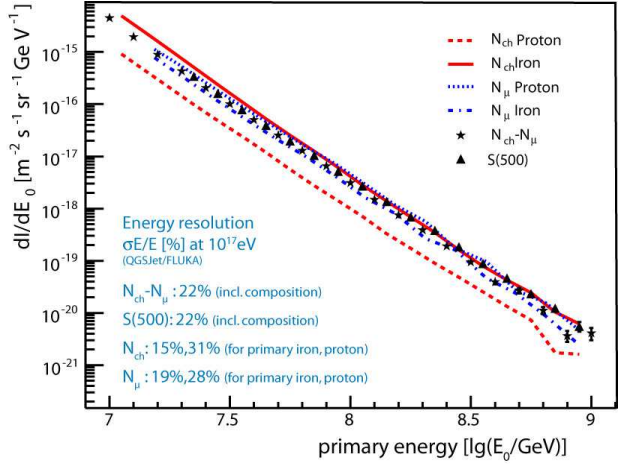}
\end{center}
\vspace*{-0.3cm}
\caption{Reconstructed all-particle energy spectrum by four different methods 
applied to KASCADE-Grande data. Given are also the energy resolutions for the methods.}
\label{fig2}
\vspace*{0.01cm}
\end{figure}
In figure~\ref{fig2} the resulting spectra are compiled. 
Owing to the different procedures, the results for the first two methods are shown 
under proton and iron assumption, respectively, only, whereas for the other 
two methods the resulting all-particle energy spectra are displayed. 
Figure~\ref{fig3} shows the same results but with the flux multiplied by a factor of $E^{3.0}$.

\subsection{Systematic uncertainties and attenuation}

The application of the different methods allows us to compare and cross-check the influence of various 
sources of systematic uncertainties. 
The $N_{ch}$-method uses the basic measurements of the Grande array only, resulting in a high 
accuracy of $N_{ch}$ with better than 15\% over the whole range of shower size, without any energy 
dependent bias. But, using only one observable, there is a large dependence on the 
primary elemental composition, reflected by the distance between the spectra obtained for 
proton and iron assumption at the calibration. 
The $N_{\mu}$-method on the other hand is based on the muon shower size, which can be 
derived less accurately (25\% with a small bias dependent on the distance of the shower 
core to the muon detectors which is corrected for), but with much less composition dependence.
The $N_{ch}$-$N_\mu$-method, owing to the combination of the reconstruction uncertainty 
of two variables, shows basically a larger uncertainty in the reconstruction, but this is compensated 
by taking into account the correlation of these observables at individual events. 
Furthermore, by this procedure the composition dependence is strikingly decreased. 
The $S(500)$ value by construction yields a larger uncertainty of the variable reconstruction, but 
has also a minor composition dependence. 
\begin{figure}[ht]
\vspace*{0.05cm}
\begin{center}
\includegraphics [width=0.49\textwidth]{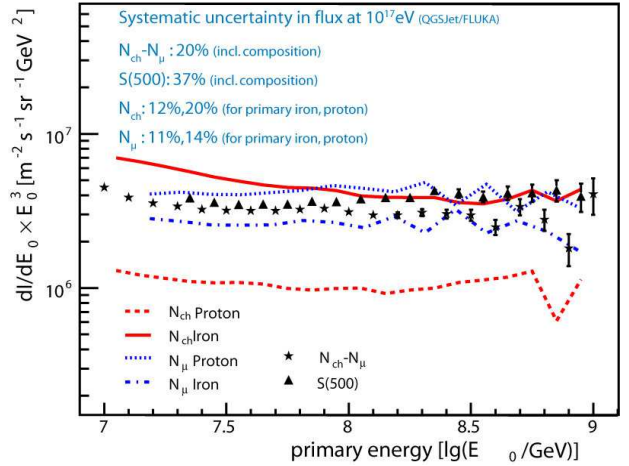}
\end{center}
\vspace*{-0.3cm}
\caption{Same as figure~\ref{fig2}, but the flux multiplied by $E_0^{3}$. Values for the uncertainty in 
the flux determination are given for the different methods.}
\label{fig3}
\vspace*{0.01cm}
\end{figure}

For all methods, the energy resolution is estimated using full Monte Carlo 
simulations and comparing the reconstructed with the simulated primary energy 
(for instance figure~\ref{fig2} gives the numbers for an energy of $E_0=10^{17}\,$eV). 
Values of systematic uncertainties in the flux determination for the different methods 
are shown in fig.~\ref{fig3} (again for $E_0=10^{17}\,$eV).
To a large amount these uncertainties are due to the reconstruction of the observables, but
there are additional sources of systematics which belong to all methods: 
e.g.,~concerning the Monte Carlo statistics, the assumed Monte Carlo spectral slope, 
or the fits of the calibration procedures.
The different attenuation (and its handling to correct for) of the various observables 
($\Lambda(N_{ch})\approx 495 \pm 20\,$g/cm$^2$; 
$\Lambda(N_{\mu})\approx 1100 \pm 100\,$g/cm$^2$; 
$\Lambda(S(500))\approx 347 \pm 22\,$g/cm$^2$ at $E_0=10^{17}\,$eV) 
however, lead again to slightly different contributions to the total systematic uncertainty. 
The total uncertainties (energy resolution and systematics) for the various methods are discussed in
refs.~\cite{kang,arteaga,bertaina,toma} and can be displayed as a band surrounding the reconstructed 
energy spectrum (e.g.,~see fig.~\ref{fig6}).  

\subsection{Discussion}

Taking into account the systematic uncertainties, there is a fair agreement between the all-particle 
energy spectra of the different applications (fig.~\ref{fig3}). 

Of particular interest is the fact that by using $N_{ch}$, the iron assumption predicts a 
higher flux than the proton assumption, whereas using $N_{\mu}$ the opposite is the case. That means 
that the `true´ spectrum has to be a solution inside the range spanned by the two methods.
If there is only the possibility of applying one method, then there is a large variance in 
possible solutions (everything within the range spanned by proton and iron line, not even 
parallel to these lines). 
However, more detailed investigations have shown, that a structure in the spectrum or a sudden change 
in composition would be retained in the resulting spectrum, even if the calibration is performed with 
an individual primary, only. 
Interestingly, over the whole energy range there is only little room for a solution satisfying both 
ranges, spanned by $N_{ch}$ and $N_{\mu}$, and this solution has to be of relatively heavy 
composition - in the framework of the QGSJet-II hadronic interaction model. 
The narrower range for a solution provided by the $N_\mu$-method compared to $N_{ch}$ 
confirms the finding of KASCADE that at sea-level the number of mostly low-energy muons 
$N_{\mu}$ is a very good and composition insensitive energy estimator.
\begin{figure}[ht]
\vspace*{0.05cm}
\begin{center}
\includegraphics [width=0.49\textwidth]{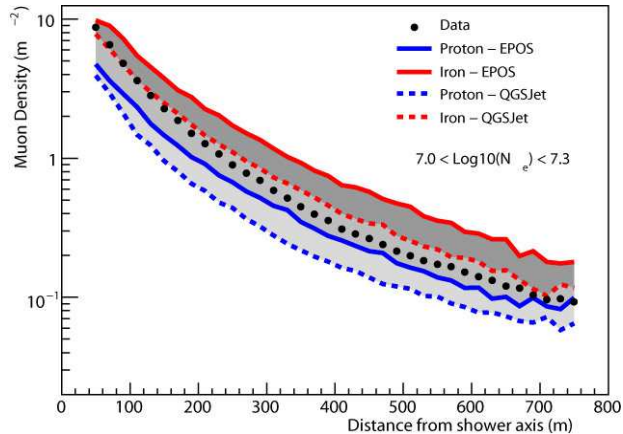}
\end{center}
\vspace*{-0.3cm}
\caption{Lateral distribution of muons compared to the predictions of
  QGSJet-II and EPOS 1.61.}
\label{figvitor1}
\vspace*{0.01cm}
\end{figure}

The results of the composition independent $N_{ch}$-$N_\mu$-, and $S(500)$-methods lie inside
the area spanned by the composition dependent methods, which is a very promising result. 
The $S(500)$-method results in a slightly higher flux than the $N_{ch}$-$N_{\mu}$-method, but 
the two spectra are consistent taking into account the systematic uncertainties. 

All the discussed results show a smooth all-particle energy spectrum without a clear hint to 
a distinct structure over the whole energy range from 10 PeV to 1 EeV. 
Another conclusion is that, taking into account the systematic uncertainties for all methods, 
the underlying hadronic interaction model (QGSJet-II/FLUKA) is intrinsically consistent, 
i.e.~the correlation between the different observables, respectively the particle components can 
describe the global features of our measurements.

\section{Hadronic interaction models}

By now, for all the considerations the models QGSJet-II and FLUKA~\cite{cors,qgs,fluka} 
have been used, only. Other interaction models would probably change the interpretation of the data. 
In the following it is briefly described how we investigate the influence of
hadronic interaction models on the interpretation of the measured data.

\subsection{Muon densities}

Local muon densities measured with the KASCADE array are studied in their
correlation with distance from the shower axis and the total
number of electrons in the shower~\cite{souza}. 
\begin{figure}[ht]
\vspace*{0.05cm}
\begin{center}
\includegraphics [width=0.49\textwidth]{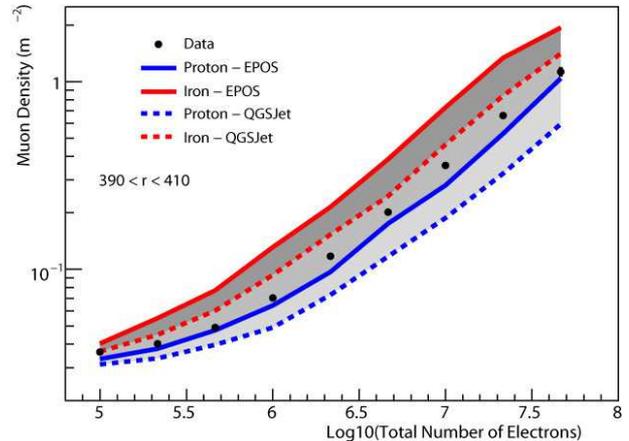}
\end{center}
\vspace*{-0.3cm}
\caption{Muon density as a function of the total number of electrons
  compared to the predictions of QGSJet-II and EPOS 1.61.}
\label{figvitor2}
\vspace*{0.01cm}
\end{figure}

Figure~\ref{figvitor1} shows the mean muon density as a function of the
distance from the shower axis compared to the predictions of QGSJet-II
and EPOS 1.61. Both hadronic interaction models include the data
within the proton and iron limits for the entire range of distances
from 100 to 750 meters. Not only the absolute density but also the slope of the 
LDF is of interest for checking the validity of hadronic interaction models. 
Considering an equal probability trigger for
protons and iron primaries as a function of distance from the shower
axis, one should expect the measured LDF to be parallel to primaries of pure 
composition. Note that the LDFs of simulated proton and iron showers are
parallel. However the measured LDF is neither parallel to the QGSJet-II nor
to the EPOS 1.61 curves. It shows that the slope of the LDF cannot
be well described by either model.

Figure~\ref{figvitor2} shows the evolution of the mean muon density
as a function of $N_e$. The calculations done with QGSJet-II
and EPOS 1.61 using  proton and iron nuclei as primary particles embrace the data in
the entire range of $5 < lg(N_{e}) < 8$.

Nevertheless, both figures~\ref{figvitor1} and~\ref{figvitor2} show that EPOS
1.61 would require a very light primary composition in order to fit the
data. On the other hand, QGSJet-II could fit the data with an intermediate
primary abundance between proton and iron nuclei.

\subsection{All-particle energy spectrum}

\begin{figure}[!t]
\vspace*{0.05cm}
\begin{center}
\includegraphics [width=0.48\textwidth]{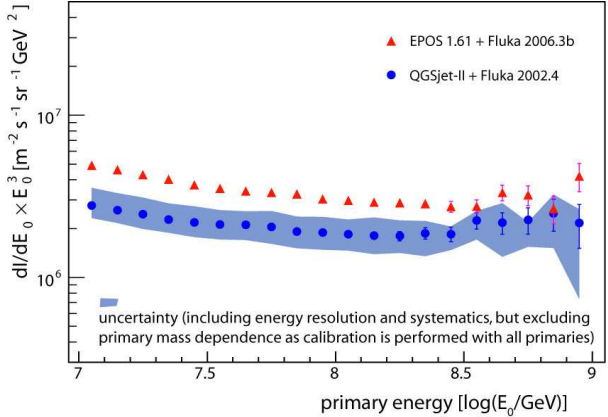}
\end{center}
\vspace*{-0.3cm}
\caption{Reconstructed all-particle energy spectrum with the $N_{ch}$-method and the calibration 
function obtained by assuming mixed composition, but based on two different 
hadronic interaction models.}
\label{fig6}
\vspace*{0.01cm}
\end{figure}

To study effects on the all-particle energy spectrum we investigated the influence 
of the hadronic interaction model exemplary by performing the 
$N_{ch}$-method based on simulations with the hadronic interaction model EPOS vers.1.61~\cite{epos}
in addition to QGSJet-II. 
As the Monte Carlo statistics is limited in case of EPOS, both spectra were obtained by generating the 
calibration curve with an equally mixed composition of five primaries (H,He,C,Si,Fe). 
Figure~\ref{fig6} compares the all-particle energy spectrum obtained with the KASCADE-Grande data set 
for both cases. 
The interpretation of the KASCADE-Grande data with EPOS leads to a significantly higher flux 
compared to the QGSJet-II result. 
Though we know, that version 1.61 of the EPOS model is not fully consistent with air shower data (in 
particular, it cannot describe the correlation of hadronic observables with the muon and electron 
content of the EAS~\cite{kasepos}) this example shows that applying and comparing various 
reconstruction methods on the same data set will be useful for a better understanding 
of the interaction processes in the air shower development.    

\subsection{Muon Tracking Detector}

More general investigations of the characteristics of the hadronic interaction models are 
provided by studies of the muon tracking data~\cite{doll09}.
Main observables of interest are the 
mean muon radial angles and the mean muon pseudo-rapidities~\cite{janusz09}, which are reconstructed 
in a given distance range from the shower core, and which are quantities sensitive to the
development of the extensive air shower.
Vertical showers ($\theta$ $\leq$ $18^{\circ}$) with size  lg$N_e$ $>$ 6 have been selected.
The lateral distributions of those two quantities have been obtained and compared with the results
reconstructed out of the simulated data for two primary species: proton and iron. The simulations were
done with CORSIKA~\cite{cors} code using the QGSJet-II model~\cite{qgs} for high energy interactions above 200 GeV and
FLUKA2006~\cite{fluka} below that energy.
The comparisons are done in limited ranges of muon distances to the core, where the saturation
effects (seen below $150\,$m) and trigger inefficiencies (seen above $400\,$m) are not present.
\begin{figure}[ht]
\vspace*{0.05cm}
\begin{center}
\includegraphics [width=0.42\textwidth]{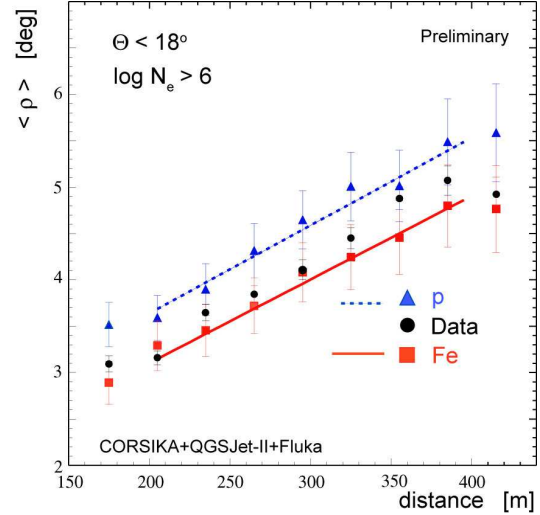}
\end{center}
\vspace*{-0.3cm}
\caption{Reconstructed lateral distribution of the mean radial angle
  compared with CORSIKA simulation results for proton and iron primaries.
  Lines are fits to the simulations.}
\label{figjan1}
\vspace*{0.01cm}
\end{figure}
\begin{figure}[ht]
\vspace*{0.05cm}
\begin{center}
\includegraphics [width=0.42\textwidth]{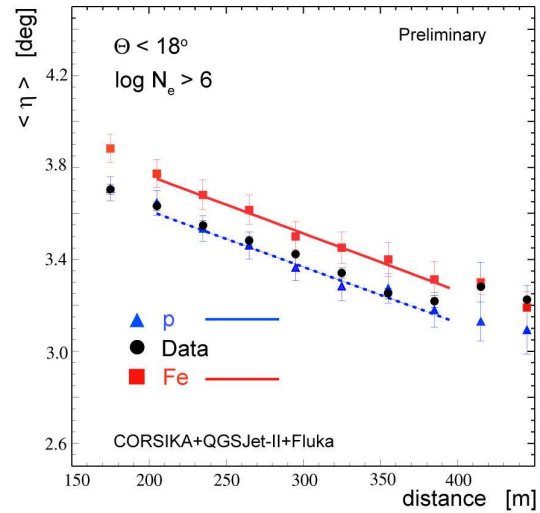}
\end{center}
\vspace*{-0.3cm}
\caption{Reconstructed lateral distribution of the mean muon pseudorapidity
  compared with CORSIKA simulation results for proton and iron primaries.
  Lines are fits to the simulations.}
\label{figjan2}
\vspace*{0.01cm}
\end{figure}
For the mean radial angle (fig.~\ref{figjan1}) we can conclude that the experimental 
data are compatible with the CORSIKA simulations using the QGSJet-II - FLUKA model combination,
as the data points are in between the simulated ones. The lines are linear fits to the simulation results. 
The error bars in the simulations are still very large and the number of simulated data will be 
increased, thus the results are marked as preliminary.
We also notice that the experimental data tend to be positioned closer to the line of
iron initiated showers rather than proton ones, like for the muon densities shown in 
fig.~\ref{figvitor1}.
In fig.~\ref{figjan2}, where lateral distributions of the mean muon pseudorapidity 
for the same data set are compared, one can also conclude that the experimental data
points are bracketed by the simulated distributions, however, the data points here are 
closer to the proton simulation results.
This inconsistency may be an indication of the features of the models. 
The mean radial angle distribution suggests that the transverse momentum of pions produced 
in hadronic interactions is reproduced by the models, but fig.~\ref{figjan2} is a hint 
that the simulations provide relatively more high rapidity pions.

\section{Composition}

The basic goal of the KASCADE-Grande experiment is the determination of the chemical 
composition in the primary energy range $10^{16} - 10^{18}\,$eV. 
Like for the reconstruction of the energy, again several methods using different observables 
will be applied to the registered data, in order to study systematic uncertainties. 
However, the influence of predictions of the hadronic interaction models has a much larger influence 
on the composition than compared to the primary energy.  
The main observables taken into account for composition studies at KASCADE-Grande are the shower size ($N_e$) 
and muon size ($N_\mu$). 
For all the methods it is crucial to verify the sensitivity of the observables to different primary particles and 
the reproducibility of the measurements with the hadronic interaction model in use as a function of sizes and 
the atmospheric depth. 
Two examples how we approach the composition studies are discussed in the following: 

\subsection{Muon densities}

In addition to the validity checks of the hadronic interaction models
the reconstruction of the local muon density at a certain distance to 
the shower core gives a sensitivity to changes in the 
elemental composition~\cite{souza} (see figs.~\ref{figvitor1} and \ref{figvitor2}).
The data as shown in figure~\ref{figvitor2} can be used to study a possible transition of     
the primary composition with increasing total number of electrons, but conclusive results are 
presently not possible due to the strong dependence on 
the hadronic interaction model. 
However, the analysis done with both models show no 
abrupt change in the composition in the entire energy range. 
EPOS 1.61 would require a very light abundance of primary particles in order to
fit the data. QGSJet-II could fit the data with an intermediate primary
abundance.
The change in slope seen in figure~\ref{figvitor2} for $log_{10}(N_e) < 6.0$
corresponds to the threshold of the experiment and the fact that both
data and simulation show the same behavior illustrates the good level
of understanding of our detector. 

\subsection{Electron-muon number ratio}

In the second example, the total number of electrons $N_e$ and the total 
number of muons $N_\mu$ of each recorded event are considered and the distribution of 
$N_\mu / N_e$ is studied in different intervals of 
$N_e$ (corresponding to different energy intervals)  and zenith angle~\cite{cantoni}.

KASCADE-Grande data are chosen, at first, in an electron size range providing full reconstruction 
efficiency and high statistics: $6.5 \leq log_{10}(N_e) < 6.75$ in $0^{\circ} \leq \theta < 24^{\circ}$
(equivalent to a primary energy just below $10^{17}\,$eV).
The same event selection is made on the simulated QGSJet-II data sets at disposal for each cosmic ray 
primary (p, He, C, Si, Fe) simulated with an energy spectrum with slope $\gamma = - 3$. 
The experimental distribution of the observable $N_\mu / N_e$ is taken into account and fitted with a 
linear combination of elemental contributions from simulations. 
By this way the mean and the width of the distributions are taken into account.
\begin{figure}[ht]
\vspace*{0.05cm}
\begin{center}
\includegraphics [width=0.49\textwidth]{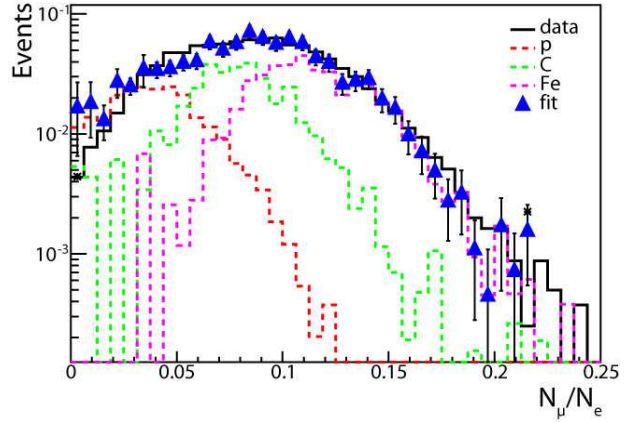}
\end{center}
\vspace*{-0.3cm}
\caption{The KASCADE-Grande data in $6.5 \leq lg(N_e) < 6.75$ ($0^\circ \leq \theta < 24^\circ$) 
described by Protons, Carbon and Iron primaries.}
\label{elena1}
\vspace*{0.01cm}
\end{figure}
\begin{figure}[h]
\vspace*{0.05cm}
\begin{center}
\includegraphics [width=0.49\textwidth]{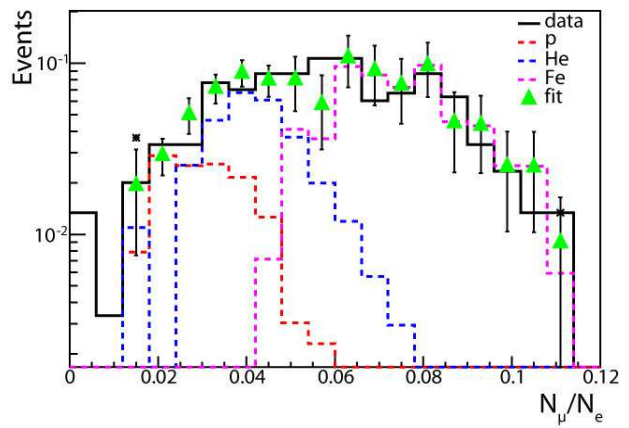}
\end{center}
\vspace*{-0.3cm}
\caption{The KASCADE-Grande data in $7.25 \leq lg(N_e) < 7.5$ ($0^\circ \leq \theta < 24^\circ$) 
described by Protons, Helium and Iron primaries.}
\label{elena2}
\vspace*{0.01cm}
\end{figure}
\begin{figure*}[!t]
\vspace*{0.05cm}
\begin{center}
\includegraphics [width=0.95\textwidth]{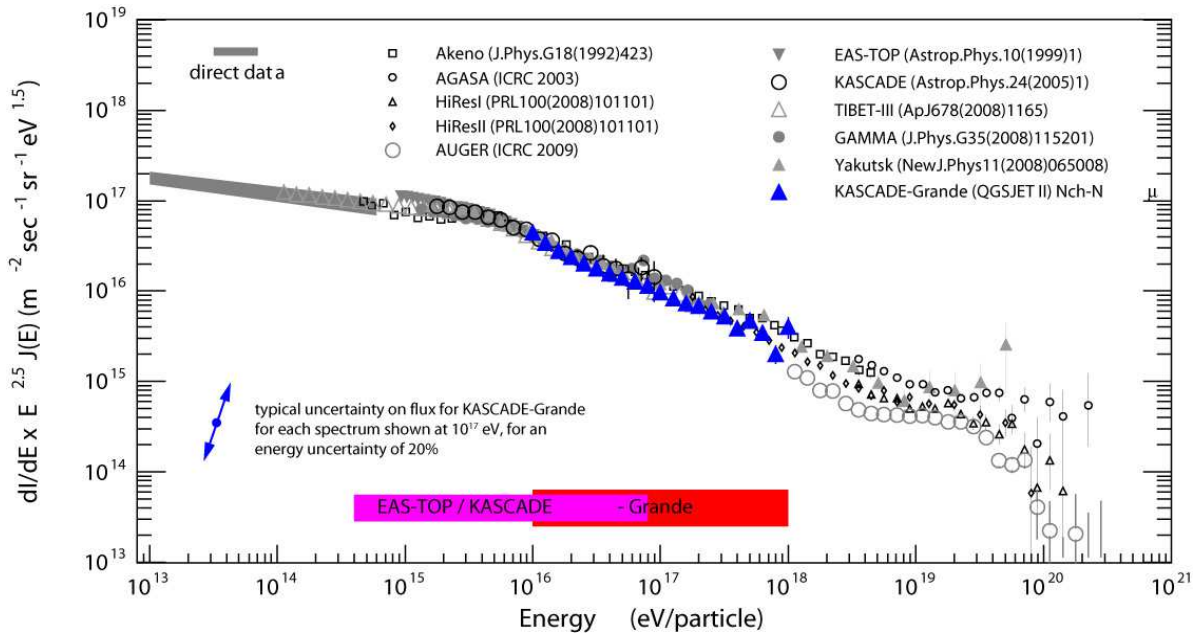}
\end{center}
\vspace*{-0.3cm}
\caption{Comparison of the all-particle energy spectrum obtained with
KASCADE-Grande data to results of other experiments.}
\label{fig7}
\end{figure*}
It was found that by using only two primary masses for the fit, the measured distributions can not be described.  
Fig.~\ref{elena1} shows the fit performed with a combination of three elements: Protons, carbon and iron which is 
matching the light elements with the heaviest element. It can be seen that a three elements combination is well 
fitting the data, but it is also possible to use another combination of primaries, e.g. protons, helium and iron. 
However, protons and iron primaries are needed to describe the edges and the width of the data distribution.
These results show that the QGSJet model describes well the shape and the tails of the experimental distribution.
Selecting the KASCADE-Grande experimental data for higher values of the electron size $N_e$ means to choose showers 
that were generated by higher energy events (fig.~\ref{elena2} is for energies slightly above $10^{17}\,$eV). 
Also in this case (as well as for other ranges in zenith angle), it is found that the model reproduces the data and the minimization of the $N_\mu / N_e$ 
distribution with three chemical components still provides a good result.

Summarizing, by such first composition studies it is seen that, by exploiting a chi-square 
minimization of a linear combination of different simulated primaries, the KASCADE-Grande N$_{\mu}$/N$_{e}$ 
distributions can only be fitted using at least three primary elements. 
QGSJet-II, the hadronic interaction model in use, can fairly well reproduce the data and, in particular, 
the tails of the distributions, that represent a main constraint being related to the lightest and heaviest 
cosmic ray primaries. 

One has to remark, that using another hadronic interaction model would of course lead to significant 
changes in the relative abundances of the elemental groups as different models predict different means. 
But, as the width of the distributions are more or less independent of the interaction model, 
the fact that at least three primary mass groups are needed to describe the data is valid.

\section{Conclusions}

Applying various different reconstruction methods to the KASCADE-Grande data  
the obtained all-particle energy spectra are compared for cross-checks of reconstruction, 
for studies of systematic uncertainties and  for testing the validity of the underlying 
hadronic interaction models. The resulting energy spectra are consistent  
and in the overlapping energy range in a very good agreement to the spectrum obtained by the 
KASCADE, EAS-TOP, and other experiments (fig.~\ref{fig7}).
A wealth of information on individual showers is available with KASCADE-Grande. This makes it possible  
to reconstruct the all-particle energy spectrum with high precision, as well as to investigate the elemental 
composition, to test the hadronic interaction models, and to study cosmic ray anisotropies. 
All these studies are under way and results are expected in near future.

\section{Acknowledgements}

The KASCADE-Grande collaboration acknowledges the possibility to present their first results
in a highlight talk at the 31$^{st}$ ICRC.
KASCADE-Grande is supported by
the BMBF of Germany, the MIUR and INAF of Italy, the
Polish Ministry of Science and Higher Education (grant for 2009-2011),
and the Romanian Ministry of Education, Research and Innovation (grant 461/2009).
This work was supported in part by the German-Polish bilateral collaboration grant
(PPP - DAAD) for the years 2009-2010.

\end{document}